\documentclass[%
 aip,
 amsmath,amssymb,
 reprint,%
]{revtex4-1}

\usepackage{graphicx}

\usepackage{color}
\usepackage{url,hyperref,lineno}
\usepackage{dcolumn}
\usepackage{bm}

\usepackage[utf8]{inputenc}
\usepackage[T1]{fontenc}
\usepackage{mathptmx}

\begin{document}

\preprint{AIP/123-QED}

\title[Mapping atmospheric waves]{Mapping atmospheric waves and unveiling  phase coherent structures in a global surface air temperature reanalysis dataset}

\author{Dario A. Zappala}
\affiliation{ 
Departament de Fisica, Universitat Politecnica de Catalunya, St. Nebridi 22, 08222 Terrassa, Barcelona, Spain
}%
\author{Marcelo Barreiro}%
\affiliation{ 
Instituto de Fisica, Facultad de Ciencias, Universidad de la
Republica, Igua 4225, Montevideo 11400, Uruguay
}%

\author{Cristina Masoller}
 \homepage{http://www.fisica.edu.uy/~cris/}
 \email{cristina.masoller@upc.edu}
\affiliation{%
Departament de Fisica, Universitat Politecnica de Catalunya, St. Nebridi 22, 08222 Terrassa, Barcelona, Spain
}%

\date{\today}

\begin{abstract}
In the analysis of empirical signals, detecting correlations that capture genuine interactions between the elements of a complex system is a challenging task with applications across disciplines. Here we analyze a global data set of surface air temperature (SAT) with daily resolution. Hilbert analysis is used to obtain phase, instantaneous frequency and amplitude information of SAT seasonal cycles in different geographical zones. The analysis of the phase dynamics reveals large regions with coherent seasonality. The analysis of the instantaneous frequencies uncovers clean wave patterns formed by alternating regions of negative and positive correlations. In contrast, the analysis of the amplitude dynamics uncovers wave patterns with additional large-scale structures. These structures are interpreted as due to the fact that the amplitude dynamics is affected by processes that act in long and short time scales, while the dynamics of the instantaneous frequency is mainly governed by fast processes. Therefore,  Hilbert analysis allows to disentangle climatic processes and to track planetary atmospheric waves. Our results are relevant for the analysis of complex oscillatory signals because they offer a general strategy for uncovering interactions that act at different time scales.\end{abstract}

\maketitle

\begin{quotation}
In our ``big data'' times, extracting useful information from complex signals is an important challenge with applications across disciplines. Due to the presence of multiple time-scales, climatological signals are particularly challenging to analyze. Here we present a technique based on the Hilbert transform that, when applied to time series of surface air temperature (with daily resolution, covering the last 30 years) unveils clear wave patterns that are interpreted as due to Rossby waves (these are atmospheric waves that propagate across our planet and have a major influence on weather). We also show that the patterns uncovered by analyzing anomaly times series include additional structures which likely appear due to climatic phenomena that have long time-scales. 
\end{quotation}

\section{Introduction}

Since the pioneer work of Lorenz~\cite{Lorenz}, the Earth system has been considered a paradigmatic case of a chaotic system and a lot of efforts have been made in the 1970s and in the 1980s to detect chaos in climate data~\cite{nicolis, grassberger, jas1986,search,kurths,tsonis}. However, the Earth system is not just chaotic but is a complex system, composed by many nonlinear subsystems with nonlinear interactions that act with different spatial and temporal scales~\cite{book_henk}. Thus, in the last two decades, the complex network paradigm~\cite{review} has been a popular framework for studying our climate (see, e.g., \cite{book}, \cite{knature} and references therein). Climate networks are constructed by using the same methodology as functional brain networks, whose links represent statistically significant similarities of signals recorded by different electrodes \cite{dante_2005,klaus_2008}. In a climate network the nodes (or vertices) are the spatial grid points of a climate dataset (time series of temperature, wind, precipitation, etc., provided on a regular grid of geographical locations), and the links (or edges) represent statistically significant similarities between pairs of time series \cite{avi_2008,tsonis_2008,donges_2009,deza_2015}. 

In both cases, to obtain meaningful networks is crucial to detect genuine interdependencies~\cite{palus_2011,hinka,giulio}. In the case of climate networks, this is particularly challenging due to the seasonality imposed by the annual solar cycle. Seasonality is typically removed by analyzing ``anomaly'' time series, instead of ``raw'' time series. Anomalies are calculated by de-trending the time series to remove the annual cycle: depending on the temporal resolution, the average hourly, daily o monthly value of the variable is subtracted. This procedure, however, does not allow to completely remove the seasonality because the fluctuations in the ``anomaly'' time series might still have a degree of seasonality (for example, large fluctuations can be common during winters and unfrequented during summers). Then, depending on the measure used to detect climatic interdependencies, statistically significant similarities can be found between the climate in distant regions, which do not disclose causal information~\cite{giulio2}. These similarities can be expected because the physical processes that govern our climate generate similar climates in distant regions.

An alternative approach for detecting genuine interdependencies that we explore in this paper is based on the analysis of ``raw'' time series (not the anomaly values) for the detection of correlated variations in the instantaneous amplitude and phase of the seasonal cycle in different geographical zones. 

\begin{figure}[!t]
\includegraphics[width=1.0\columnwidth]{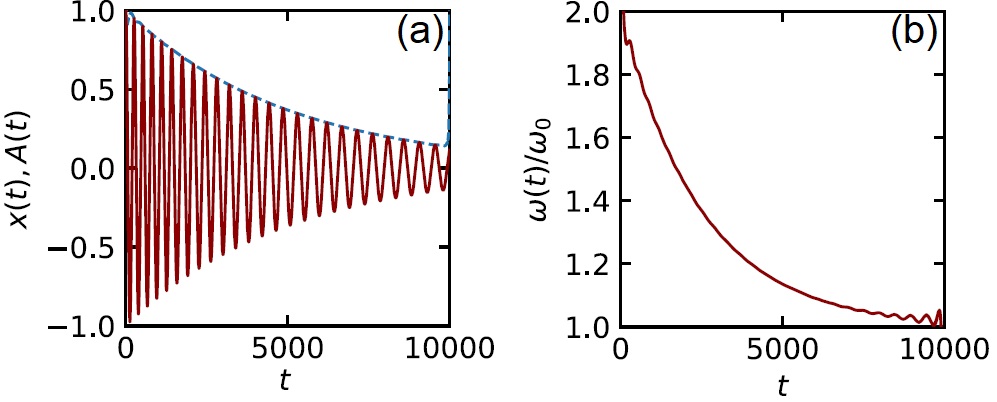}\\
\caption{Hilbert amplitude and instantaneous frequency computed from the time series of an oscillatory variable with time-varying amplitude and frequency, $x(t)=e^{-\alpha t}\cos[\omega_0(1+e^{-2\alpha t})]$, where $\omega_0=2\pi/400$, $\alpha=2/T$ and $T=10^5$ is the length of the time series. (a) $x(t)$ (solid line) and the Hilbert amplitude (dashed line); (b) normalized instantaneous frequency. As expected (except near the extremes) the Hilbert amplitude and the instantaneous frequency exponentially decay in time. Examples of the analysis of real SAT signals can be found in  ~\cite{dario1,dario2,dario3}. \label{fig1}}
\end{figure}

There are several ways to define, for each data point of a real oscillatory signal, $x(t)$, an instantaneous amplitude and phase~\cite{pikbook}. A popular one is to remove the mean value and represent the signal on the complex plane, $s(t)=x(t)+i x^H(t)$, where $x^H(t)$ is the Hilbert transform (HT) of $x(t)$. Then, writing $s(t)$ in polar coordinates, $s(t)=a(t)\exp[i\phi(t)]$, defines the instantaneous amplitude, $a(t)$, and the instantaneous phase, $\phi(t)$. The derivative of the unwrapped phase (not restricted to the interval [$-\pi,\pi$]) is the instantaneous angular frequency, $\omega(t)$ (an example is presented in Fig.~\ref{fig1}).

While this procedure can be applied to any real signal, the amplitude, the phase and the instantaneous frequency have a clear physical meaning of rotation only if the signal is sufficiently narrow band. In this case, the amplitude is the envelope of the oscillation and the time-averaged instantaneous frequency coincides with the frequency of the dominant peak in the Fourier spectrum~\cite{pikbook}. However, the HT has also been used for the analysis of empirical signals which do not fulfill the narrow-band condition (e.g., electric brain activity~\cite{mario,tass_1998}). In order to obtain narrow-band behavior, these signals are often pre-filtered in the frequency domain (by selecting a narrow frequency band of interest), or in the temporal domain (by using a sliding averaging window). 

We argue that the HT directly applied to raw broadband oscillatory signals can yield valuable information. In Refs.~\cite{dario1,dario2,dario3} we have analyzed a global data set of surface air temperature (SAT) with daily resolution and we have shown that the HT yields meaningful phase and amplitude information of the seasonal cycles in different geographical zones. A video demonstrating the seasonal variation of the phase of SAT (i.e., disregarding amplitude information), can be found in~\cite{video}. Analyzing the dynamics of the instantaneous frequency in the different regions we have uncovered large-scale spatial regions of high average frequency~\cite{dario1}. Contrasting the mean values of $\left<a(t)\right>_t$ and $\left<\omega(t)\right>_t$ calculated in different decades we have identified the geographical regions where the variations were most pronounced~\cite{dario2}. Averaging over time using sliding windows and changing the window size, we were able to unveiled hidden periodicities in the data~\cite{dario3}. 

While in~\cite{dario1,dario2,dario3} we performed univariate analysis (i.e., we analyzed the properties of $a_i(t)$ and $\omega_i(t)$ in single geographical locations, here we perform bivariate analysis: we analyze the cross-correlation of $a_i(t)$, $a_j(t)$ (and of $\omega_i(t)$, $\omega_j(t)$) for pairs of geographical locations. Our goal is to detect similarities in the temporal dynamics that can reveal genuine interdependencies and causal interactions. We also search for phase coherence (or phase coherent seasonality) that can be masked by amplitude fluctuations. Using the Kuramoto parameter~\cite{kuramoto1,kuramoto2} as a measure of phase coherence we find strong asymmetries in the north and in the south hemispheres. We also find that the correlations detected with the instantaneous frequency time series are the ones that provide the most clear signature of Rossby waves. Rossby waves are giant meanders in high-altitude winds that are due to the rotation of the planet \cite{book_henk}. They are important sources of atmospheric variability and their identification and characterization (temporal duration, spatial extension)~\cite{caro,avi,sholomo} can advance weather predictability in the north and in the south hemispheres~\cite{RWs}.  Our results demonstrate that the Hilbert method allows to disentangle climatic processes and to track the surface temperature signature of Rossby waves. We also show that the wave structure, formed by alternating regions of negative and positive correlations, becomes distorted when SAT is pre-filtered in the temporal domain by using a sliding averaging window of several days.

This paper is organized as follows. Section 2 describes the dataset analyzed and method used, Sec. 3 presents the results, and Sec. 4 presents the conclusions.

\section{Dataset and Methods}
\label{Data}

We use the Era-Interim reanalysis that is a globally gridded dataset that represents the state variables of the climate system, incorporating observations and the output of a  weather prediction model~\cite{ERAdatos}. The data analyzed covers the period from 1979 to 2017, has daily resolution and spatial resolution of 2.5$^\circ$ in latitude and in longitude ($73 \times 144 $ grid points). Therefore, we have $N= 10512$ SAT time series, each with $L = 14245$ data points (i.e., days). To indicate the raw SAT time series we use the notation $r_j(t)$, where $j \in [1, N]$ represents the geographical site and $t \in [1, L]$ represents the day.
	
For each time series we first remove the linear trend and normalize the time series to zero mean and unit variance. The detrended and normalized time series is referred to as $x_j(t)$. The Hilbert transform of $x_j(t)$, $HT[x_j](t)=y_j(t)$, allows defining an analytic signal, $h_j(t) = x_j(t) + i y_j(t)$, from which we calculate the amplitude and phase time series as $a_j(t)=\sqrt{x_j^2(t) + y^2_j(t)}$, $\varphi_j(t) = \arctan [{y_j(t)}/{x_j(t)}]$ with $\varphi_j(t)$ in $[-\pi, \pi]$. Then, we unwrap the phase series and calculate the instantaneous frequency time series as the time
derivative of the unwrapped phase. 
	
Because the HT calculated over a finite time gives errors near the extremes of the series {(an example is presented in Fig.~\ref{fig1})}, we disregard the initial and final two years~\cite{dario2,dario3}. Thus, we analyze time series of 35 years with $12783$ data points.

We use the Kuramoto parameter \cite{kuramoto1,kuramoto2} to measure the degree of phase coherence in a geographical region $S$. Because the sites have different areas, we compute the area-weighted Kuramoto parameter as
\begin{equation}
r(t) = \left | \frac{\sum_{j\in S}w_j \exp[i\varphi_j(t)]}{\sum_{j\in S} w_j} \right |.
\end{equation}
Here $w_j=\cos(\lambda_j)$ is proportional to the area of site $j$, with $\lambda_j$ being the latitude of site $j$.

\begin{figure}[!t]
\includegraphics[width=.5\textwidth]{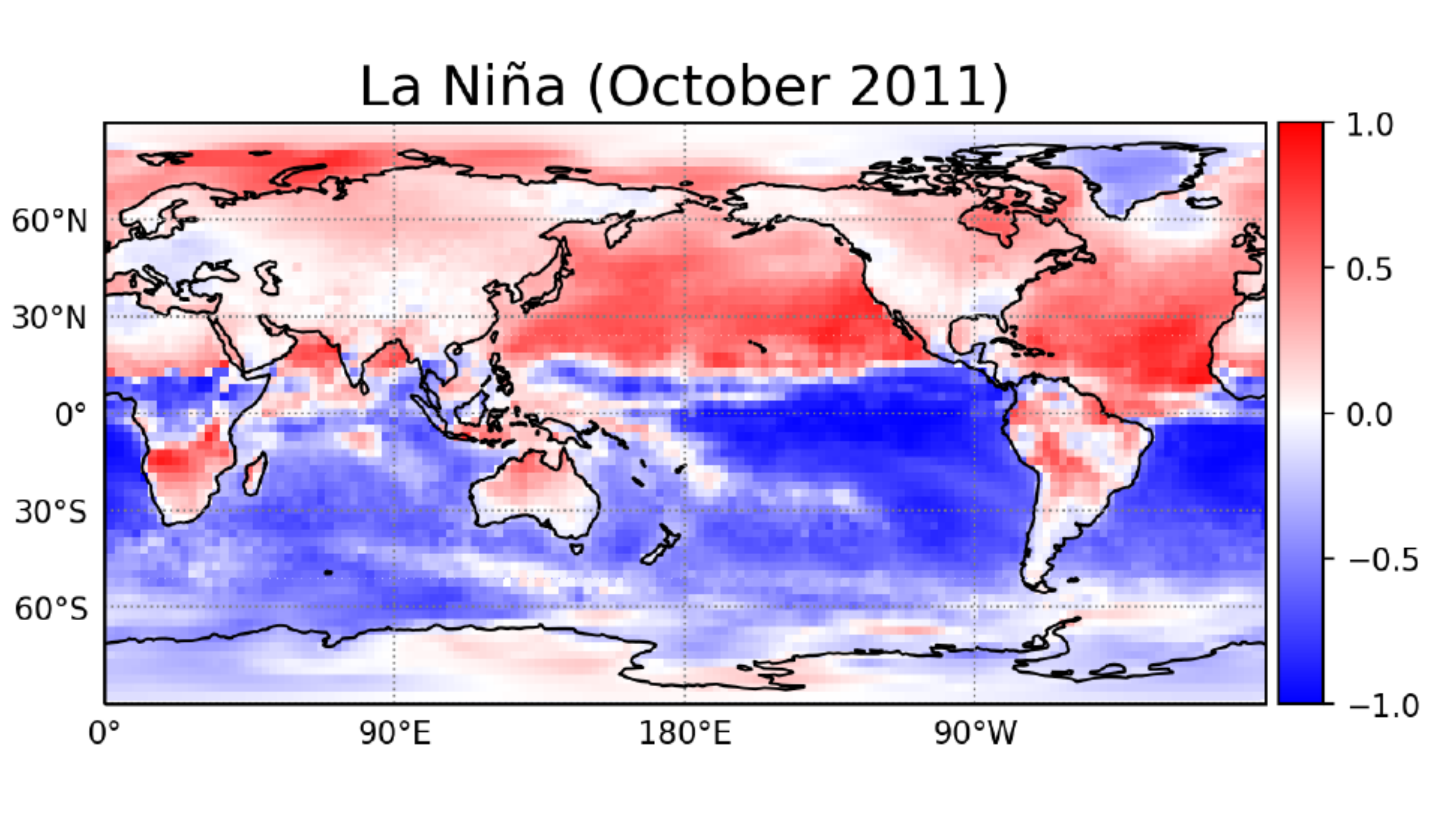}\\%
\includegraphics[width=.5\textwidth]{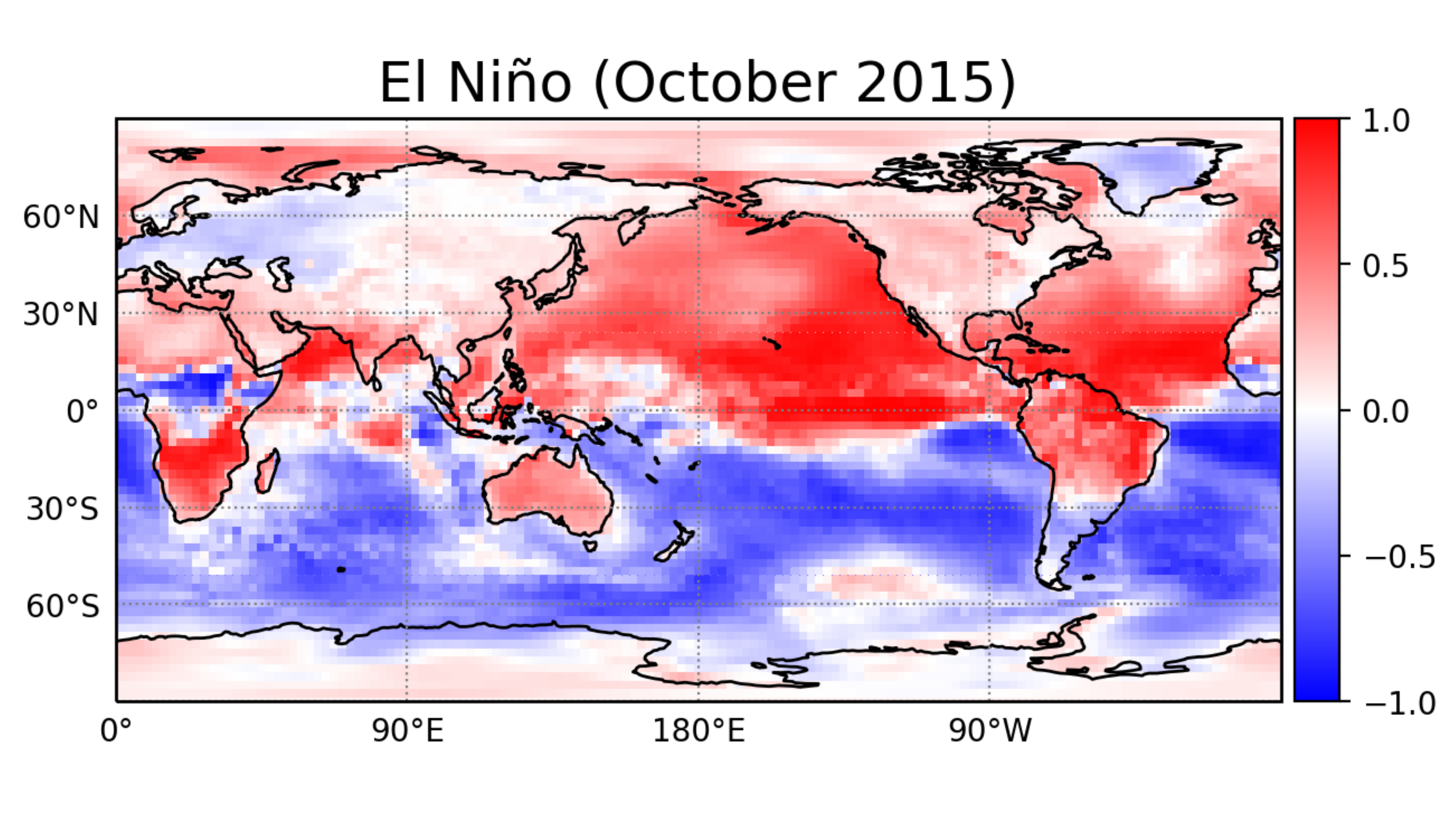}
\caption{Cosine of the Hilbert phase, averaged over one month during La Ni\~na period and during a El Ni\~no period. Due to the annual seasonal cycle, the two hemispheres are in antiphase (October is autumn in the north and is spring in the south). In  La Ni\~na period, the sea surface temperature (SST) in the eastern tropical Pacific region decreases, cooling down the atmosphere, which is seen in the map as a blue region; in contrast, during El Ni\~no period, SST increases and heats the atmosphere, which is seen in the map as a red region. The cosine of the Hilbert phase also uncovers differences during La Ni\~na and El Ni\~no periods in other regions of the world (Australia, South Africa and South America). A video demonstrating the seasonal variation of the phase of SAT can be found in~\cite{video}.\label{fig2}}
\end{figure}

\section{Results}
\label{Results}

To give a first insight of SAT phase dynamics, we present in Fig.~\ref{fig2} a snapshot of the cosine of the phase (in color code, averaged over one month),  during a period of La Ni\~na (October 2011, top panel) and El Ni\~no (October 2015, bottom panel).
La Ni\~na are El Ni\~no are two stages of the same phenomenon, the El Ni\~no–Southern Oscillation (ENSO). In El Ni\~no stage, the sea surface temperature (SST) of the  eastern tropical Pacific increases by a few degrees, and this in turn heats the  atmosphere. This is seen in the cosine of the phase as a red region in the eastern tropical Pacific region. In contrast, in La Ni\~na stage, SST decreases, which cools down the atmosphere, which is seen in the map as a blue region. We note that, as expected, the two hemispheres are in antiphase (October is autumn in the north and is spring in the south). 

\begin{figure}[!t]
\includegraphics[width=1.0\columnwidth]{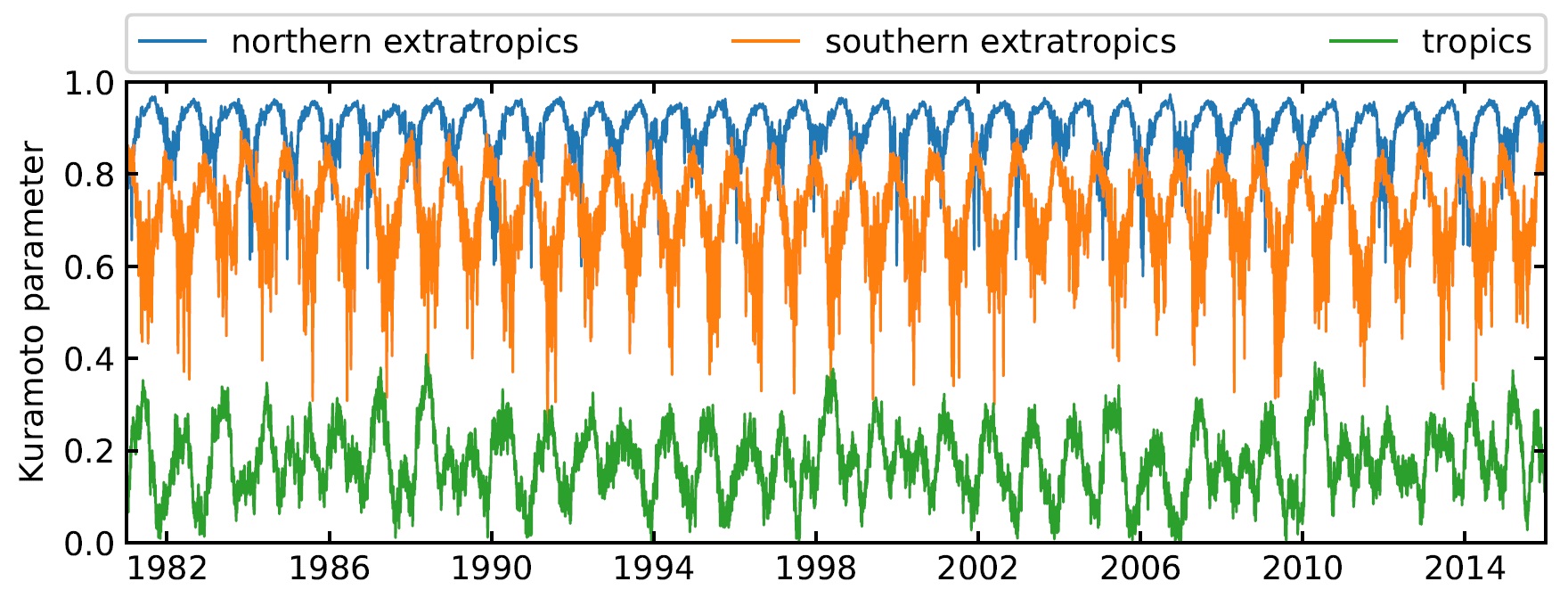}
\caption{Temporal evolution of the area-weighted Kuramoto parameter, Eq. (1), in the northern extratropics (north of 30$^\circ$), in the southern extratropics (south of -30$^\circ$) and in the tropical region (in between -30$^\circ$ and 30$^\circ$). The Kuramoto parameter reveals a higher degree of phase coherence in the northern extratropics with respect to the southern extratropics, and almost no coherence in the tropics. We also note a seasonal variation of the level of phase coherence, with higher coherence during the summers.\label{fig3}}
\end{figure}

To measure the degree of phase coherence we compute the Kuramoto parameter, Eq. (1), in three large geographical areas: in the northern extratropics (north of 30$^\circ$), in the southern extratropics (south of -30$^\circ$) and in the tropical belt (in between -30$^\circ$ and 30$^\circ$). The results are presented in Fig.~\ref{fig3}. We see high level of phase coherence in the northern extratropics, a lower level in the southern extratropics, and almost no phase coherence in the tropics. We also note a seasonal variation in the {{level of phase coherence}}, with an oscillation that is wider in the southern extratropics, in comparison to the northern extratropics. Also, in both hemispheres, phase coherence is higher during summers than during winters.

\begin{figure*}[!t]
\includegraphics[width=.95\textwidth]{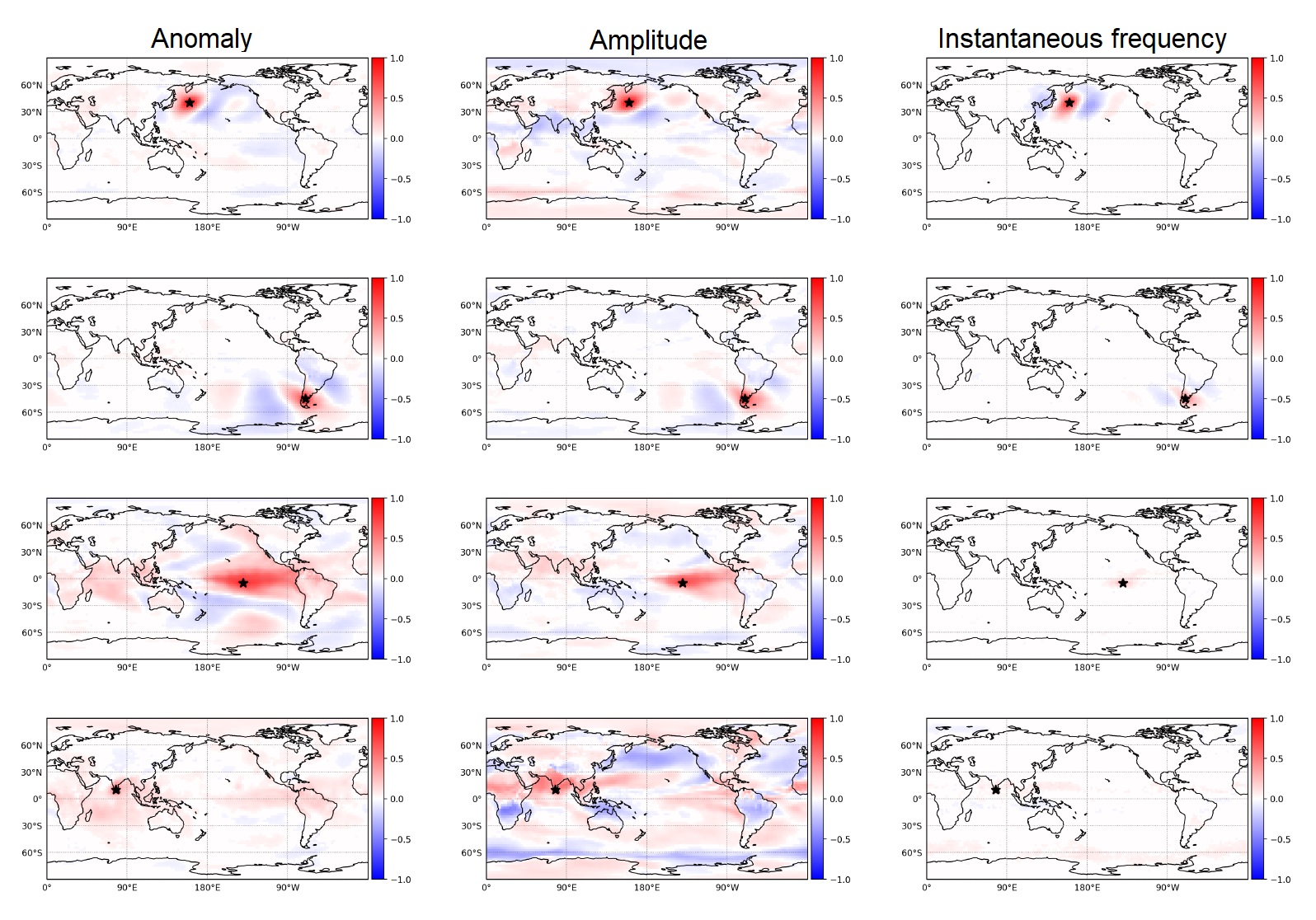}
\caption{Cross-correlation between a time series in a particular region (indicated with a symbol) and the same time series in all other regions. From top to bottom, the selected region is in the northern extratropics (40N, 160E), in the southern extratropics (45S, 70W), in El Ni\~no area (5S, 140W) and in India (10N, 77.5E). The columns represent correlations computed from the anomaly time series (left), the amplitude time series (center), and the  instantaneous frequency time series (right). We note that the frequency maps in the extratropical sites have clear wave structure as they are composed only by bands of alternating signs; in contrast, the maps obtained from the amplitude or from the anomaly time series have no clear wave structure. For both tropical sites the frequency cross-correlation is almost zero, likely due to the fact that in the tropics atmospheric dynamics is governed by fast convective phenomena with small spatial coherence.\label{fig4}}
\end{figure*}

To investigate similarities in the time series of the Hilbert amplitude and the instantaneous frequency, we select four sites (in the northern extratropics, in the southern extratropics, in India, and in the central Pacific Ocean) and analyze how their times series are correlated with other regions of the world. For comparison, we also calculate the correlation of the anomaly time series. 

We begin by analyzing zero-lag correlations. We calculate the linear cross correlation coefficient, $\left <x_i(t)x_j(t) \right >_t$, with $x_i(t)$ and $x_j(t)$ representing either the Hilbert amplitude, or the instantaneous frequency, or the anomaly time series at sites $i$ and $j$, normalized to zero-mean and unit variance. 

The results are displayed in Fig.~\ref{fig4}.  Anomaly (left column) gives strong correlation patterns in the two extratropical sites, with bands of opposite signs, and also in the El Niño site, while it gives weaker correlations in the Indian site. The Hilbert amplitude (central column) gives strong correlation patterns in all four sites. We note that in the two extratropical sites,  the maps obtained from anomaly and from amplitude are qualitatively similar. In contrast, the instantaneous frequency (right column) reveals almost no correlation in the two tropical sites, while it yields, in the extratropical sites, patterns that are partially similar to those given by anomaly, albeit with a more clear wave structure. These results can be interpreted in the following terms. The tropical atmosphere can not sustain large-scale horizontal temperature gradients, and convection has small horizontal scales; in the extratropics the presence of Rossby waves induces in the surface low and high pressure systems, whose associated warm and cold fronts generate the  alternating positive and negative surface temperature correlations.

Taken together the maps in both extratropical sites, we note that the patterns uncovered by instantaneous frequency correlations give the most clear signature of Rossby waves, as they are formed only by bands of alternating signs.

We have compared these maps with those obtained using the Spearman coefficient (a nonlinear correlation measure, not shown) and found  identical spatial structures.

To further verify that temporal correlations in the dynamics of the instantaneous frequency uncover patterns that are characteristic of Rossby waves, we calculated the lagged cross-correlation coefficient, $\left <x_i(t)x_j(t+l) \right >_t$, with $l$ in days. In Fig.~\ref{fig6} we see that, as $l$ increases, the wave pattern moves, as expected, towards east. We have also studied the effect of pre-filtering the fast SAT variability before applying the Hilbert transform (averaging SAT in a temporal window of several days). The results, shown in Fig.~\ref{fig6}, reveal a distortion of the wave pattern (as additional structures emerge) and a change in the wave number (that counts the number of times the wave fits in a circle at a given latitude). These modifications are consistent with the fact that the temporal average filters out synoptic fluctuations. In the synoptic scale, Rossby waves have, in the southern hemisphere, a zonal wave number that varies in the range 4-11 but has a maximum in between 5 and 6. Low-frequency waves (periods larger than 10 days, which are uncovered when filtering out the synoptic fluctuations) tend to have a lower zonal wave number \cite{Trenberth}. For future work, it will be interesting to use the Hilbert approach to further investigate the results reported in Ref.~\cite{fernando}, that connectivity patterns due to stationary Rossby waves differ during El Ni\~no and La Ni\~na: during  El Ni\~no the wave train originates close to Australia, while during La Ni\~na, it seems to be  generated from the central Pacific.

\begin{figure}[!t]
\includegraphics[width=.3\textwidth]{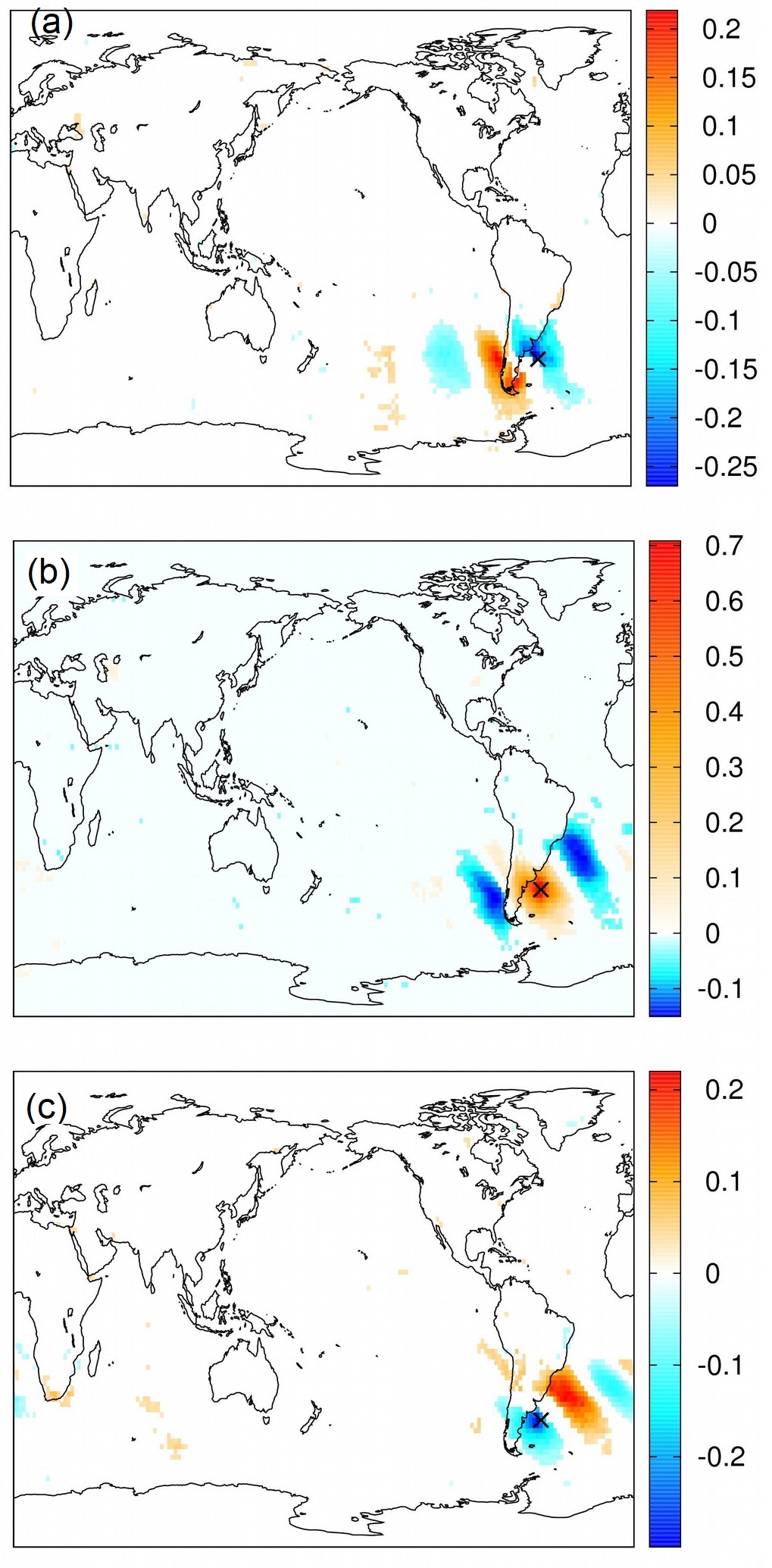}
\caption{Cross-correlation map {{computed from the instantaneous frequencies, considering a}} site in southern extratropics (indicated with a symbol) and using a lag equal to (a) -2, (b) 0, (c) 2 days. The color scale is different in the different panels, for a clear visualization of the wave pattern. We observe that with increasing lag the wave pattern moves towards the east.\label{fig5}}
\end{figure}

\begin{figure}[!t]
\includegraphics[width=.4\textwidth]{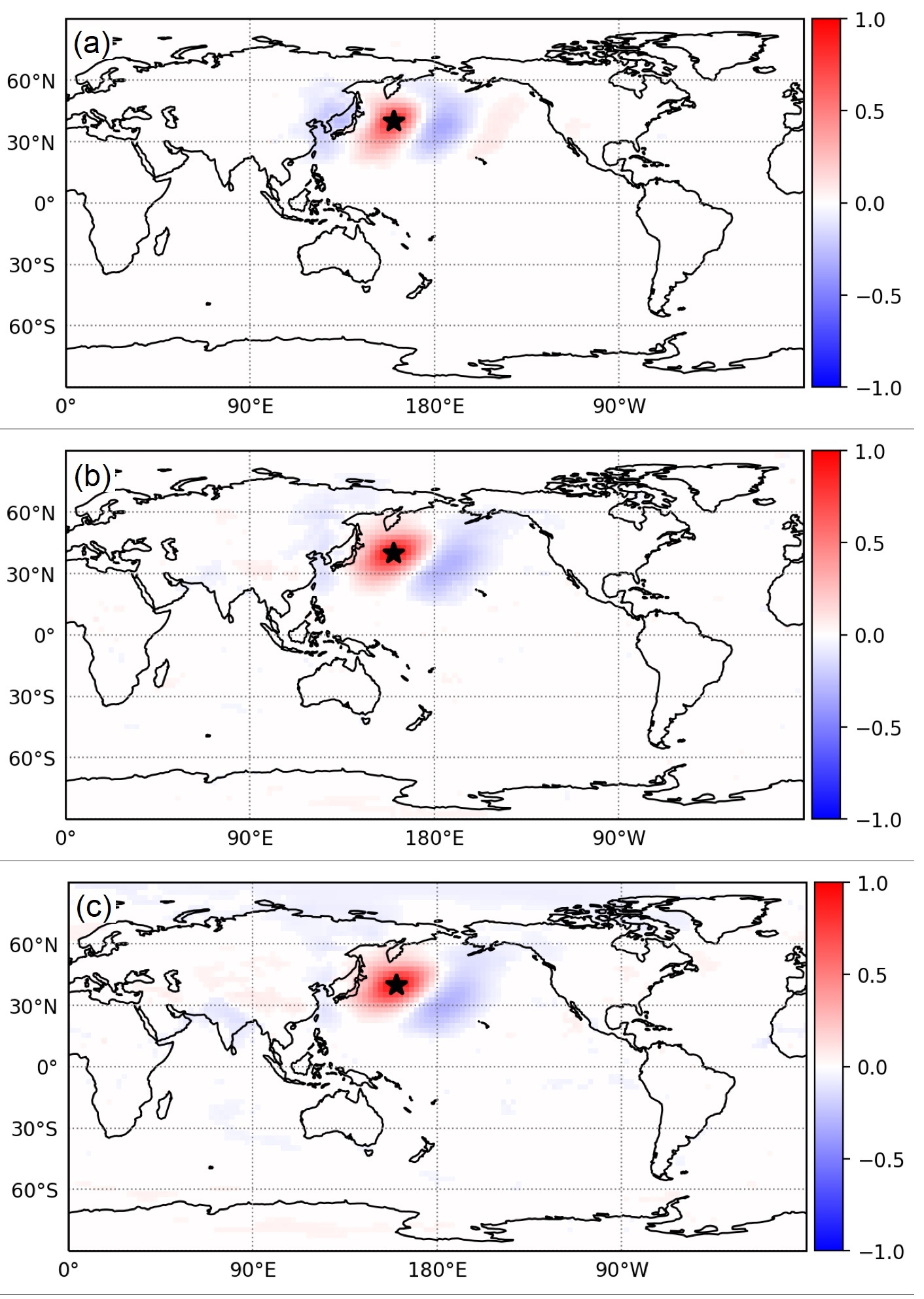}
\caption{Influence of averaging the SAT time series in a sliding window, before applying the Hilbert transform. The color code displays the cross-correlation computed from the instantaneous frequencies, considering a site in northern extratropics (indicated with a symbol), when SAT is averaged in a window of (a) 0, (b) 15, (c) 31 days. We note that averaging SAT produces a distortion of the wave pattern, as the bands of alternating signs increase in size and additional structures appear.\label{fig6}}
\end{figure}

\section{Conclusions}\label{Conclusions}

To summarize, we have used Hilbert analysis to investigate phase coherent seasonality and correlations in SAT time series. We calculated the temporal evolution of the area-weighted Kuramoto parameter in three main regions: the northern extratropics, the southern extratropics, and the tropical region. We have found that both extratropics have a high degree of phase coherence, with the northern extratropics being more coherent than the southern extratropics. In contrast, the tropical region has a low level of phase coherence. In the three regions the Kuramoto parameter has seasonal regularity and in both, the southern and the northern extratropics, phase coherence is strongest in the summer.  The Kuramoto parameter reveals different degrees of seasonality in the different hemispheres, as it has larger fluctuations in the southern extratropics. The higher phase coherence in the northern extratropics can be explained by the presence of larger landmasses that give an annual cycle of SAT with a wider amplitude. Another factor that can account for higher phase coherence in the northern extratropics is the presence in the atmosphere of standing Rossby waves. These are generated by the interaction between atmospheric circulation and topography, as well as by land-sea temperature contrasts, and both mechanisms are more important in the northern hemisphere than in the southern hemisphere. 

We have also analyzed the cross-correlation maps obtained from anomaly, amplitude and instantaneous frequency time series. We found that, in the extratropics, the instantaneous frequency gives the most clean patterns formed by bands of alternating signs, which are a clear surface signature of upper-level Rossby waves. Further evidence of Rossby waves was found when computing the cross-correlation of the instantaneous frequencies with a lag time. 
Overall, our work yields new insights into the detection of correlations that are not partially due to similar seasonality, which is a problem for constructing climate networks whose links capture genuine interactions. It also offers a new approach to track the surface signatures of Rossby waves. Moreover, our work is relevant to the wide scientific community working in complex systems because we have shown that the analysis of the instantaneous frequency extracted from oscillatory signals, without pre-filtering the data to smooth out temporal fluctuations, can yield meaningful information.

\begin{acknowledgments}
This work was supported in part by Spanish Ministerio de Ciencia, Innovación y Universidades  (PGC2018-099443-B-I00), AGAUR FI scholarship (D. Z.) and ICREA ACADEMIA (C. M.), Generalitat de Catalunya.
\end{acknowledgments}


\end{document}